\documentclass[prl,twocolumn,showpacs,twoside,10pt,superscriptaddress]{revtex4}
\usepackage{amssymb}

\usepackage{amsmath}
\usepackage{graphicx}

\input{tcilatex}

\begin{document}

\title{Experimental realization of quantum games on a quantum computer}
\author{Jiangfeng Du}
\email{djf@ustc.edu.cn}
\affiliation{Department of Modern Physics, University of Science and Technology of China,
Hefei, 230027, P.R.China.}
\altaffiliation{Temporary address: Service de Physique Th\'{e}orique CP225, Universit\'{e}
Libre de Bruxelles, 1050 Brussels,Belgium.}
\author{Hui Li}
\affiliation{Department of Modern Physics, University of Science and Technology of China,
Hefei, 230027, P.R.China.}
\author{Xiaodong Xu}
\affiliation{Department of Modern Physics, University of Science and Technology of China,
Hefei, 230027, P.R.China.}
\author{Mingjun Shi}
\affiliation{Department of Modern Physics, University of Science and Technology of China,
Hefei, 230027, P.R.China.}
\author{Jihui Wu}
\affiliation{Laboratory of Structure of Biology, University of Science and Technology of
China, Hefei, 230027, P.R.China.}
\author{Xianyi Zhou}
\affiliation{Department of Modern Physics, University of Science and Technology of China,
Hefei, 230027, P.R.China.}
\author{Rongdian Han}
\affiliation{Department of Modern Physics, University of Science and Technology of China,
Hefei, 230027, P.R.China.}

\begin{abstract}
We generalize the quantum Prisoner's Dilemma to the case where the players
share a non maximally entangled states. We show that the game exhibits an
intriguing structure as a function of the amount of entanglement with two
thresholds which separate a classical region, an intermediate region and a
fully quantum region. Furthermore this quantum game is experimentally
realized on our nuclear magnetic resonance quantum computer.
\end{abstract}

\maketitle

In 1982, Feynman\cite{1} observed that quantum mechanical systems have an
information-processing capability much greater than that of classical
systems, and could thus potentially be used to implement a new type of
powerful computer. Three years later Deutsch\cite{2} described a
quantum-mechanical Turing machine, showing that quantum computers could
indeed be constructed. Although the theory is well understood, actually
building a quantum computer has proved extremely difficult. Up to now, only
three methods have been used to demonstrate quantum logical gates: Trapped
ions\cite{3}, cavity QED\cite{4} and NMR\cite{5}. Of these methods, NMR has
been the most successful with realizations of quantum teleportation\cite{6},
quantum error correction\cite{7}, quantum simulation\cite{8}, quantum
algorithms\cite{9} and others\cite{10}. In this Letter, we add game theory%
\cite{11} to the list: quantum games can be experimentally realized on a
nuclear magnetic resonance quantum computer.

Recently a new application of quantum information to game theory has been
discovered\cite{12,13,14,15,16,17}. Game theory is an important branch of
applied mathematics. It is the theory of decision-making and conflict
between different agents. Since the seminal book of Von Neumann and
Morgenstern\cite{18}, modern game theory has found applications ranging from
economics through biology\cite{19,20}. In the process of a game, whenever a
player passes his decision to other players or the game's arbiter, he
communicates information. Therefore it is natural to consider the
generalization when the information is quantum, rather than classical\cite%
{12,13}. It should also be noted that many problems in quantum information
theory can be considered as quantum games, for instance quantum cloning\cite%
{21}, quantum cryptography\cite{22} and quantum algorithms\cite{13}.

The Prisoner's Dilemma is a famous game in classical game theory and has
been extended into quantum domain by Eisert \textit{et al.}\cite{12}. Their
work was based on the maximally entangled state. In this Letter, we
generalize the quantum Prisoner's Dilemma to the case where the players
share a non maximally entangled states. We show that the game exhibits an
intriguing structure as a function of the amount of entanglement. In
addition we have realized this quantum game on our nuclear magnetic
resonance quantum computer. We believe that it is the first explicit
physical realization of such a quantum game.

\begin{figure}[tbp]
\includegraphics{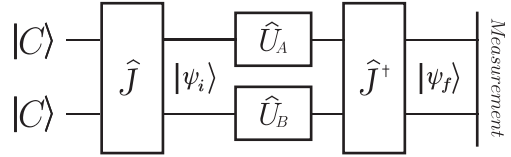}
\caption{The setup for the two-player quantum game.}
\label{Fig1}
\end{figure}

\begin{table}[b]
\caption{Payoff matrix for the Prisoner's Dilemma. The first entry in the
parenthesis denotes the payoff of Alice and the second of Bob.}%
\begin{ruledtabular}
\begin{tabular}{ccc}
& Bob: $\hat{C}$ & Bob: $\hat{D}$ \\ \hline
Alice: $\hat{C}$ & $\left( \alpha ,\beta \right) $ & $\left( \gamma ,\gamma
\right) $ \\
Alice: $\hat{D}$ & $\left( \gamma ,\gamma \right) $ & $\left( \beta ,\alpha
\right) $%
\end{tabular}
\end{ruledtabular}
\label{Table1}
\end{table}

Let us now briefly recall the quantum Prisoner's Dilemma presented in Ref%
\cite{12}. There are 2 players, the players have 2 possible strategies:
cooperate($\widehat{C}$) and defect($\hat{D}$). The payoff table for the
players is shown in Table \ref{Table1}. Classically the dominant strategy
for both players is to defect(the Nash Equilibrium) since no player can
improve his/her payoff by unilaterally changing his own strategy, even
though the \textit{Pareto optimal }is for both players to cooperate. This is
the dilemma. In the quantum version, see Fig.\ref{Fig1}, one starts with the
product state $\left| C\right\rangle \left| C\right\rangle $. One then acts
on the state with the entangling gate $\widehat{J}$ to obtain $\left| \psi
_{i}\right\rangle =\widehat{J}\left| CC\right\rangle =1/\sqrt{2}\left(
\left| CC\right\rangle +i\left| DD\right\rangle \right) $. The players now
act with a local unitary operator $\hat{U}_{A}$ and $\hat{U}_{B}$\ on their
qubit. Finally the disentangling gate $\widehat{J}^{+}$ is carried out and
the system is measured in the computational basis, giving rise to one of the
four outcome $\left| CC\right\rangle $,$\left| CD\right\rangle $,$\left|
DC\right\rangle $,$\left| DD\right\rangle $. If $\hat{U}_{A}$ and $\hat{U}%
_{B}$ are restricted to the classical strategy space $\left( \widehat{C}=%
\widehat{I},\widehat{D}=i\widehat{\sigma }_{y}\right) $, then one recovers
the classical game. If one allows quantum strategies of the form

\begin{equation}
\hat{U}\left( \theta ,\phi \right) =\left( 
\begin{array}{cc}
e^{i\phi }\cos \theta /2 & \sin \theta /2 \\ 
-\sin \theta /2 & e^{-i\phi }\cos \theta /2%
\end{array}%
\right)
\end{equation}%
with $0\leqslant \theta \leqslant \pi $ and $0\leqslant \phi \leqslant \pi
/2 $, then there exists a new Nash Equilibrium, label $\hat{Q}\otimes \hat{Q}
$, with the payoff $\left( 3,3\right) $. It has the property of being 
\textit{Pareto optimal}, therefore the dilemma that exists in the classical
game is resolved. It was pointed out in Ref\cite{12} that if one allows any
local operations, then there is no longer a unique Nash Equilibrium.

\begin{figure}[tbp]
\includegraphics{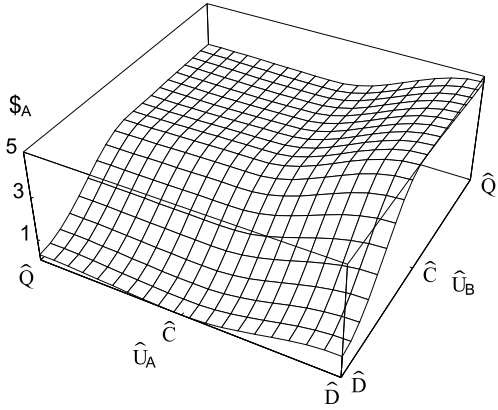}
\caption{Alice's payoff for $\protect\gamma =\protect\gamma _{th1}/2$. In
this and the following two plots, we have chosen a parametrization such that
the strategies $\hat{U}_{A} $\ and $\hat{U}_{B}$\ each depend on a single
parameter $t\in \left[ -1,1\right] $: $\hat{U}_{A}=\hat{U}\left( t\protect%
\pi ,0\right) $\ for $t\in \left[ 0,1\right] $\ and $\hat{U}_{A}=\hat{U}%
\left( 0,-t\protect\pi /2\right) $\ for $t\in \left[ -1,0\right] $ (same for
Bob). Cooperation $ \hat{C}$\ corresponds to the value $t=0$, defection $%
\hat{D}$\ to $t=1$, and $\hat{Q}$\ to $t=-1$.}
\label{Fig2}
\end{figure}

In the present letter we generalize Eisert et.al.'s scheme by taking the
entangling operation to have the form $\left| \psi _{i}\right\rangle =%
\widehat{J}\left| CC\right\rangle =\cos (\gamma /2)\left| CC\right\rangle
+i\sin (\gamma /2)\left| DD\right\rangle $, where $\gamma \in \lbrack 0,\pi
/2]$ measures the entanglement of the initial state. We shall restrict
ourselves to strategies of the form of eq(1). We will show that an
intriguing structure emerges as $\gamma $ is varied from $0$ (no
entanglement) to $\pi /2$\ (maximally entanglement), namely the game has two
thresholds, $\gamma _{th1}=\arcsin \sqrt{1/5}$ and $\gamma _{th2}=\arcsin 
\sqrt{2/5}$. Fig.\ref{Fig2} indicates Alice's expected payoff for $\gamma =$ 
$\gamma _{th1}/2$.

\begin{figure}[tbp]
\includegraphics{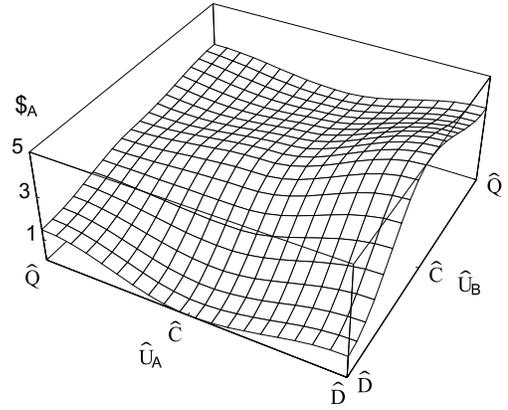}
\caption{Alice's payoff for $\protect\gamma =\left( \protect\gamma _{th1}+%
\protect\gamma _{th2}\right) /2$. The parametrization is chosen as in Fig.%
\ref{Fig2}.}
\label{Fig3}
\end{figure}

\begin{figure}[tbp]
\includegraphics{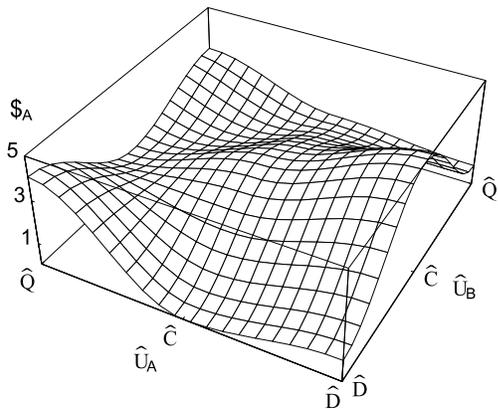}
\caption{Alice's payoff for $\protect\gamma =\left( \protect\gamma _{th2}+%
\protect\pi /2\right) /2$. The parametrization is the same as Fig.\ref{Fig2}.
}
\label{Fig4}
\end{figure}

In this case the game has features similar to the separable game with $%
\gamma =0$, see Ref\cite{12}. Indeed for $0\leqslant \gamma \leqslant \gamma
_{th1},$ the quantum game behaves ``classically'', \textit{i.e.} the only
Nash Equilibrium is $\hat{D}\otimes \hat{D}$ and the payoffs for the players
are both $1$, which is the same as in the classical game. Fig.\ref{Fig3}
shows Alice's expected payoff with $\gamma =$ $(\gamma _{th1}+\gamma
_{th2})/2$. Assuming Bob chooses $\hat{D}=\hat{U}(\pi ,0)$, Alice's best
strategy is $\hat{Q}=\hat{U}(0,\pi /2)$ with $\$_{A}(\hat{Q},\hat{D})=5\sin
^{2}\gamma $; while assuming Bob's strategy is $\hat{Q}$, Alice's optimal
reply is $\hat{D} $ with $\$_{A}(\hat{D},\hat{Q})=5\cos ^{2}\gamma $. Since
the game is symmetric, the same holds for Bob. Thus, $\hat{D}\otimes \hat{D}$
is no longer a Nash Equilibrium because each player can improve his/her
payoff by unilaterally deviating from the strategy $\hat{D}$. However, two
new Nash equilibria $\hat{Q}\otimes \hat{D}$ and $\hat{D}\otimes \hat{Q}$
appear. This feature holds for $\gamma _{th1}<\gamma <\gamma _{th2}$.
Indeed, $\$_{A}(\hat{U}(\theta ,\phi ),\hat{D})=\sin ^{2}(\theta /2)+5\cos
^{2}(\theta /2)\sin ^{2}\phi \sin ^{2}\gamma $ and $\$_{A}(\hat{U}(\theta
,\phi ),\hat{Q})=4-\cos \theta +(-3+2\cos \theta -\cos ^{2}(\theta /2)\cos
2\phi )\sin ^{2}\gamma $, hence $\$_{A}(\hat{U}(\theta ,\phi ),\hat{D}%
)\leqslant 5\sin ^{2}\gamma =\$_{A}(\hat{Q},\hat{D})$ and $\$_{A}(\hat{U}%
(\theta ,\phi ),\hat{Q})\leqslant 5\cos ^{2}\gamma =\$_{A}(\hat{D},\hat{Q})$
for all $\theta \in \lbrack 0,\pi ]$ and $\phi \in \lbrack 0,\pi /2]$.
Analogously $\$_{B}(\hat{D},\hat{U}_{B})\leqslant \$_{B}(\hat{D},\hat{Q}%
)=5\sin ^{2}\gamma $ and $\$_{B}(\hat{Q},\hat{U}_{B})\leqslant \$_{B}(\hat{Q}%
,\hat{D})=5\cos ^{2}\gamma $ for all $\hat{U}_{B}$. So $\hat{D}\otimes \hat{Q%
}$ and $\hat{Q}\otimes \hat{D}$ are both Nash Equilibria, with the feature
that the Payoff of the player who adopts strategy $\hat{D}$ is better than
that of the player who adopts $\hat{Q}$. Thus in this regime the quantum
game does not resolve the dilemma. But for $\gamma >\gamma _{th2}$ quantum
strategies resolve the dilemma. In Fig.\ref{Fig4} we depict Alice's payoff
as a function of the strategies $\widehat{U}_{A}$ and $\widehat{U}_{B}$ with 
$\gamma =(\gamma _{th2}+\pi /2)/2$. This figure is similar to the one for
the maximally entangled game in Ref\cite{12}. It can be shown that $\hat{Q}%
\otimes \hat{Q}$ is a unique equilibrium not only for $\gamma =(\gamma
_{th2}+\pi /2)/2$\ but also for any $\gamma \in \lbrack \gamma _{th2},\pi
/2] $. Hence a novel Nash Equilibrium $\hat{Q}\otimes \hat{Q}$ arises with
payoff $\$_{A}(\hat{Q},\hat{Q})=\$_{B}(\hat{Q},\hat{Q})=3$, which has the
property of being \textit{Pareto optimal}\cite{19}. The dilemma that exists
in the classical game is removed as long as the game's entanglement exceeds
the threshold $\gamma _{th2}=\arcsin \sqrt{2/5}\approx 0.685$, even though
the game's initial state is not maximally entangled.

\begin{figure}[tbp]
\includegraphics{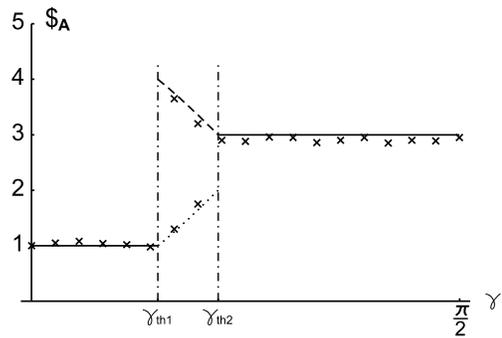}
\caption{The expected payoff for Alice as a function of the measure of the
parameter $\protect\gamma $\ when both players resort to Nash Equilibrium.
The line correponds to theoretic calculation and the cross to the
experimetal results. For $\protect\gamma _{th1}<\protect\gamma <\protect%
\gamma _{th2}$, the dashed line and the doted line represent Alice's payoff
when the Nash Equilibrium is $\hat{D}\otimes \hat{Q}$\ and $\hat{Q}\otimes 
\hat{D}$\ repectively.}
\label{Fig5}
\end{figure}

Fig.\ref{Fig5} indicates Alice's payoff as a function of the parameter $%
\gamma $ when both players resort to the Nash Equilibrium. The two
thresholds are analogous to phase transitions. When the amount of
entanglement is less than the smaller threshold, one is in a classical
region. When the amount of entanglement lies between the two thresholds, one
is in a transition region between classical and quantum behavior. The last
domain is the fully quantum region. It is surprising that in the transition
region, both Nash Equilibria result in an unfair game, even though the
structure of the game is symmetric with respect to the interchange of the
two players. We think that the reasons for the asymmetry are: (i) Since the
definition of Nash Equilibrium allows multiple Nash Equilibria to coexist,
the solutions may be degenerated. Therefore the definition itself allows the
possibility of such an asymmetry. This situation is similar to the
spontaneous symmetry breaking; (ii) If we consider the two Nash Equilibria
as a whole, they are fully equivalent and the game remains symmetric. But
finally, the two players have to choose one from the two equilibria. This
also causes the asymmetry of the game.

\begin{figure}[tbp]
\includegraphics{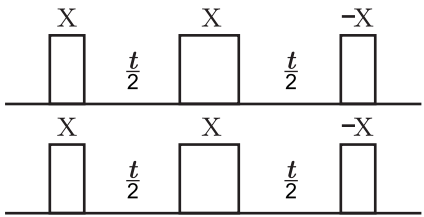}
\caption{NMR pulse sequece used to implement the entangling gate $J$. Narrow
boxes correspond to $90^{o}$\ pulses, whereas wide boxes are $180^{o}$\
pulses; the upper and lower lines refer to the nuclear spins corresponding
to Alice's and Bob's qubits, respectively; the phase of each pulse is
written above it.}
\label{Fig6}
\end{figure}

This quantum game was implemented using our two qubit NMR quantum computer,
described in Ref.\cite{23}. This computer uses the two spin states of $^{1}H$
nuclei of partially deuterated cytosine in a magnetic field as qubits, while
radio frequency (RF) fields and spin--spin couplings between the nuclei $%
J_{AB}=7.17Hz$ are used to implement quantum logic gates. Experimentally, we
performed nineteen separate sets of experiments with the entanglement of the
player's qubits given by $\gamma =n\cdot \pi /36\quad (n=\{0,1,2,\cdots
,18\})$. The $\gamma =0$ $(n=0)$ corresponds to Eisert \textit{et al.}'s
separable game and $\gamma =\pi /2$ $(n=18)$ corresponds to their maximally
entangled quantum game. In each set, the full process of the quantum game
shown in Fig.\ref{Fig1} was executed. The details of the process are as
follows: (1) The quantum game starts with the computer in the unentangled
pure state $\left| CC\right\rangle $, but with an NMR quantum computer it is
impossible to begin in a true pure state. Using the methods of Cory \textit{%
et al.}\cite{24} it is, however, possible to create an effective pure state,
which behaves in an equivalent manner. (2) The initial entangled state is
obtained by applying the entangling gate $\hat{J}=\exp \{i\gamma \hat{D}%
\otimes \hat{D}/2\}$ which was performed with the pulse sequence shown in
Fig.\ref{Fig6}, where the time period $t=\gamma /(\pi J_{AB})$.

(3) Players Alice and Bob execute their strategic moves (the Nash
equilibrium) described as local unitary operations $\widehat{U}_{A}\otimes 
\widehat{U}_{B}$. As shown above, $\widehat{U}_{A}\otimes \widehat{U}_{B}$
is determined by the value of $\gamma =\pi Jt=n\cdot \pi /36$.
Experimentally, $\widehat{D}\otimes \widehat{D}$ $(0\leqslant \gamma <\gamma
_{th1},i.e.$ $n=\{0,1,2,3,4,5\})$ was implemented using a non-selective $%
180_{y}^{o}$ pulse; $\widehat{D}\otimes \widehat{Q}$ $(\hat{Q}\otimes \hat{D}%
)$ $(\gamma _{th1}\leqslant \gamma \leqslant \gamma _{th2},i.e.$ $n=\{6,7\})$%
\ was implemented by performing a selective $180_{y}^{o}$ pulse on Alice's
(Bob's) qubit, while a selective pulse sandwich $%
90_{-y}^{o}-180_{x}^{o}-90_{y}^{o}$ was performed on Bob's (Alice's) qubit;
and $\hat{Q}\otimes \hat{Q}$ $(\gamma _{th2}\leqslant \gamma \leqslant \pi
/2,i.e.$ $n=\{8,9,\cdots ,18\})$\ was implemented using a composite
non-selective pulse sandwich $90_{-y}^{o}-180_{x}^{o}-90_{y}^{o}$. (4)
Finally, the disentangling gate $\widehat{J}^{+}=\exp \{-i\gamma \hat{D}%
\otimes \hat{D}/2\}$ (the inverse of $\widehat{J}$) is applied before the
measurement. The pulse sequence to implement $\widehat{J}^{+}$ is the same
as in Fig.\ref{Fig6} except for $t=(2\pi -\gamma )/(\pi J_{AB})$. Thus the
final state $\left| \psi _{f}\right\rangle =\left| \psi _{f}(\widehat{U}_{A},%
\widehat{U}_{B})\right\rangle $ of the game prior to measure is given by $%
\left| \psi _{f}\right\rangle =\widehat{J}^{+}(\widehat{U}_{A}\otimes 
\widehat{U}_{B})\widehat{J}\left| CC\right\rangle $.

In NMR experiment, it is not practical to determine the final state
directly, but an equivalent measurement can be made by so-called quantum
state tomography\cite{5}. The readout procedure consists of applying a
sequence of RF pulses, measure the resulting induction signal, Fourier
transform to get the spectra, and integrate to get the areas of the
resonance peaks. By applying nine different pulse sequences (no rotation,
rotation about $\widehat{x}$, and about $\widehat{y}$ , for each of the
spins), the elements in the density matrix were sampled, allowing a
least-squares procedure to recover the density matrix $\rho $ from the data.
Then the expected payoff was determined using the numerical values of the
payoff table of Prisoner's Dilemma by the $\$_{A}=3P_{CC}+5P_{DC}+P_{DD}$
and $\$_{B}=3P_{CC}+5P_{CD}+P_{DD}$, where $P_{\sigma \sigma ^{^{\prime
}}}=\left\langle \sigma \sigma ^{\prime }\right| \rho \left| \sigma \sigma
^{\prime }\right\rangle \ $\ is the probability of finding the eigenstate $%
\left| \sigma \sigma ^{\prime }\right\rangle $\ (with $\tsum\nolimits_{%
\sigma ,\sigma ^{\prime }\in \left\{ C,D\right\} }P_{\sigma \sigma ^{\prime
}}=1$).

All experiments were conducted at room temperature and pressure on Bruker
Avance DMX-500 spectrometer in Laboratory of Structure Biology, University
of Science and Technology of China. Alice's payoffs as a function of the
parameter $\gamma $\ (the measure of entanglement) in our NMR experiments
are shown in Fig.\ref{Fig5}. The computations shown in Fig.\ref{Fig1} took
less than $300$ milliseconds, which was well within the the decoherence time 
$T_{2}\approx 3s$. The relationship between player's payoff and the
parameter $\gamma $ in the quantum game is clearly seen in Fig.\ref{Fig5},
with good agreement between theory and experiment. The relative error is
less than $8\%$. The errors are primarily due to inhomogeneity of magnetic
field, imperfect $90^{o}$ and $180^{o}$ pulses, and the variability over
time of the measurement process.

In summary, it was shown in Ref.\cite{12} that the classical Prisoner's
Dilemma can be generalized into a quantum game, and that when a maximally
entangled state is employed the dilemma disappears. We used the same
physical model as Eisert et al, but introduced a new parameter $\gamma $,
which measures the amount of entanglement in the quantum game. As $\gamma $
varies, novel features appear: there are two thresholds, $\gamma _{th1}$ and$%
\ \gamma _{th2}$, which separate the classical region, an intermediate
region where 2 Nash Equilibrium coexist, and a fully quantum region where
the dilemma disappears. The fact that the dilemma can be removed as long as
the game's entanglement exceeds a certain threshold $\gamma _{th2}$, is very
much as in quantum cryptography and computation, where the superior
performance of the quantum system depends strongly on the amount of
entanglement. Furthermore, we realized this scheme experimentally on our
two-qubit ensemble quantum computer. These experimental results demonstrate
how a NMR quantum computer can load an initial state, enable each player to
perform his/her quantum strategic moves, and readout the payoffs. This
reveals a new domain of application for quantum computers.

Note added: Since this work was carried out we have generalized it in three
ways: first we have considered the correlations between entanglement and
quantum games for different sets of strategies\cite{17}, second we have
considered three-player entanglement enhanced quantum games\cite{25} and
finally we analyzed how the thresholds $\gamma _{th1}$ ,$\ \gamma _{th2}$\
vary when the parameters in the payoff table are changed\cite{26}.

We thank J. Eisert, J.W. Pan and Y. D. Zhang for helpful discussion and S.
Massar for a carefully reading of the manuscript. This work was supported by
National Nature Science Foundation of China (Grants. Nos. 10075041 and No.
10075044) and Science Foundation of USTC for young scientists.

\end{document}